# Achievement of highly radiating plasma in negative triangularity and effect of reactor-relevant seeded impurities on confinement and transport

L. Casali[1], D. Eldon[2], T. Odstrcil[2], R. Mattes[1], A. Welsh[1], K. Lee[1], A. O. Nelson[3], C. Paz-Soldan[3], F. Khabanov[4], T. Cote[2], A. G. McLean[5], F. Scotti[5], K.E. Thome[2]

Email: lcasali@utk.edu

[1]University of Tennessee-Knoxville, Knoxville, Tennessee, USA
[2]General Atomics, San Diego, CA, USA
[3]Columbia University, New York, NY, USA
[4]University of Wisconsin–Madison, Madison, WI, USA
[5]Lawrence Livermore National Laboratory, Livermore, CA, USA

**Abstract**

The first achievement of highly radiating plasmas in negative triangularity is shown with an operational space featuring high core radiation at high Greenwald fraction obtained with the injection of reactor-relevant seeded gases. These negative triangularity (NT) shape diverted discharges reach high values of normalized plasma pressure ($\beta_N >$ 2) at high radiation fraction with no ELMs. We demonstrate that as long as the impurity level in the core is kept low to avoid excessive fuel dilution and impurity accumulation, integration of NT configuration with high radiation fraction not only is achievable but it can lead to confinement improvement with stabilization effects originating from collisionality, E×B shear and profiles changes due to impurity radiation cooling. The underlying physics mechanism is robust and holds for a variety of impurity species. The absence of the requirement to stay in H-mode translates in a higher core radiation fraction potentially allowed in NT shape effectively mitigating the power exhaust issue. The results presented here demonstrate a path to high performance, ELM free and highly radiative regime with rector-relevant seeding gases making this regime a potential new scenario for reactor operation.

## 1. Introduction

A grand challenge for the tokamak approach to fusion energy production is the simultaneous achievement of a high-performance core with a boundary solution compatible with the requirements dictated by the first wall materials. Successful integration must include a boundary solution characterized by low divertor temperature < 5-10 eV to suppress physical erosion, low particle flux to reduce chemical erosion, low steady-state heat load on the divertor target $q_{perp} <$ 10MW/m$^2$ to avoid material melting and sublimation, and very limited levels of intermittent heat flux [1]. Boundary solutions rely on radiative power exhaust through injection of low and medium Z impurities in the plasma to convert the heat flux into electromagnetic radiation and redistribute it over the whole plasma vessel surface [2, 3, 4, 5, 6].

Impurity radiation is the main driver of the energy dissipation process and in addition to direct power dissipation, a major task of seeded impurities is to promote divertor detachment [7, 8]. It has been demonstrated in several tokamaks that while power exhaust and divertor detachment obtained with solely gas puff can reduce plasma performance, the injection of external radiators can lead to a simultaneous reduction of power fluxes and improved core performance [2, 5, 8, 9, 10, 11, 12, 13].

The core and the plasma boundary are governed by different physics making the problem complex because it features multi-scales and multi-species especially when impurity seeding is used. Understanding how to simultaneously fulfill the competing needs of the plasma core on one side and the divertor on the other side is critical for the design and operation of future reactors.

The standard tokamak operation mode is the H-mode regime obtained with a strongly D shaped positive triangularity (PT) [14]. The H-mode regime provides such a high-performance fusion core due to the build-up of an edge transport barrier leading to a pedestal structure compared to the low confinement mode "L-mode". However, fusion reactor design based on H-mode operation in PT faces the serious challenge of power handling: in order to operate in H-mode, the power crossing the separatrix has to exceed the L-H power threshold.

This core requirement translates to high power flux into the plasma boundary that needs to be quickly dissipated in the SOL and divertor to not damage the Plasma Facing Components (PFCs). The H-mode regime is also characterized by transient heat fluxes due to the Edge Localized Modes (ELMs) that eject an abrupt amount of energy in a very short time [15] and cannot be tolerated in future rectors.

Furthermore, the e-folding length of the SOL power flux in H-mode is driving the major divertor heat exhaust challenge in H-mode reactor studies [16, 17]. This calls for alternative plasma configurations and scenarios optimized for power handling and integrated solutions in addition to the ones in the standard D-shape scenario. Negative triangularity (NT) [18, 19, 20, 21, 22, 23, 24] has the potential to solve the issues encountered by PT by achieving H-mode confinement level with a plasma edge that resembles L-mode confinement with a moderate pedestal and therefore does not exhibit ELMs. Additionally, a NT shape naturally puts the divertor at large major radius, which means a larger wetted area for power loads is available for enhanced heat flux spreading. Hence, NT with the divertor positioned at large R/low magnetic field ($B_T$) has great potential to improve core-edge integration and easing the divertor installation, remote access and maintenance technology. The absence of the H-mode level pedestal in NT removes ELMs and the constrains of operating above the L-H power threshold while maintain good confinement. The removal of the L-H power threshold condition opens the possibility to operate in a strongly radiative mantle scenario where a significant amount of the exhaust is radiated inside the separatrix. This way the SOL will have less power density and higher particle density both promoting divertor detachment. Such conditions in NT plasmas were not explored anywhere before the results presented here.

This work reports on the first achievement of strongly shaped NT plasmas with a single lower X-point and a highly radiating mantle sustained at high confinement. In Section 2 we present the first characterization of the NT operational space at high radiation fraction obtained with the injection of extrinsic impurities that are feasible for reactor applications. The power balance analysis is discussed in Section 3 with the profile prediction by TGYRO framework investigated in Section 4. The linear stability analysis is presented in Section 5. Discussion and Conclusion are given in Section 6. The numerical tools employed in this work to interpret experimental data and to model plasma the plasma behavior are: the TRANSP code [25] for power balance analysis, the TGYRO [26] code to obtain steady-state kinetics profiles matching the power balance fluxes from TRANSP analysis, the Trapped-Gyro-Landau-Fluid (TGLF) [27] code for linear stability analysis, the NEO [28] code, a multi species drift-kinetic neoclassical solver for neoclassical calculations and the AURORA transport code [29] for the modeling of the impurity transport. In this paper, we focus on the effects of impurity seeding on plasma confinement and transport. We will present the impact of impurity seeding on SOL and divertor behavior in another publication [30].

## 2. Impurity seeding experiment

High performance, high power, vertically stable, highly radiative and robustly ELM free discharges have been achieved in negative triangularity (NT) shape diverted discharges (with negative upper and lower triangularity) enabled by the installation of a graphite-tile armor on the low-field-side lower outer wall on DIII-D [31, 32, 33, 34]. In the experiments that are the basis of this paper, we explore the NT setup with a radiating mantle obtained with a reactor-relevant seeding species. Radiative experiments were performed in feedback control and enabled by an upgrade of the radiated power control system using the core radiation as the control variable to assess the limits to high radiation fraction in NT [35]. This is an important development since independent control of both core and divertor radiation with different seeding impurities is necessary for future reactors.

The comprehensive dataset spans a wide parameter range with a total radiated power fraction up to 0.85 and a Greenwald density fraction between 0.2 and 1. Mantle radiation limits have been explored and achieved with neon, argon and krypton seeding. Nitrogen seeding has been added in some discharges as extrinsic divertor radiator in addition to carbon (which is an intrinsic impurity at DIII-D) to trigger access to detachment conditions. Key parameters scans include line average density (3e19-9e19 m$^{-3}$), plasma current (Ip= 0.8-1.0 MA), toroidal magnetic field direction ($B_T$ = -2T or 2T), heating power from neutral beam injection $P_{NBI}$ and from the maximum available electron cyclotron heating ($P_{NBI}$ = 3-5 MW, and $P_{ECH}$ = 0-2 MW), impurity seeding species (Ne, Ar, Kr, Kr+N, Ar+N), impurity seeding rate (0, 3e20 particles s$^{-1}$) and impurity seeding locations as summarized in fig.1. The signals are averaged over 100ms and the core radiation fraction $f_{rad,core}$ is defined here as everything within the Last Closed Flux Surface (LCFS) obtained with tomographic inversion with PyTomo [36]. Reactor relevant value of normalized plasma pressure $\beta_N$ ($\beta_N$ > 2) were achieved with krypton and argon at high radiation fraction and high Greenwald fraction (fig.1 right plot). Figure 2 shows a clear trend of $\beta_N$ increasing with core radiation fraction obtained with Kr and Ar seeding while confinement degradation is observed at high $f_{rad,core}$ for neon due to high dilution. The highest $\beta_N$ values at around $f_{rad,tot}$ ~ 0.75 correspond to a discharge where argon seeding rapidly leads to a 25% increase in $\beta_N$ with ~30% increase in radiation. The results presented here indicate a path to high performance, ELM free and highly radiative regime with rector-relevant seeding gases in a negative triangularity configuration.

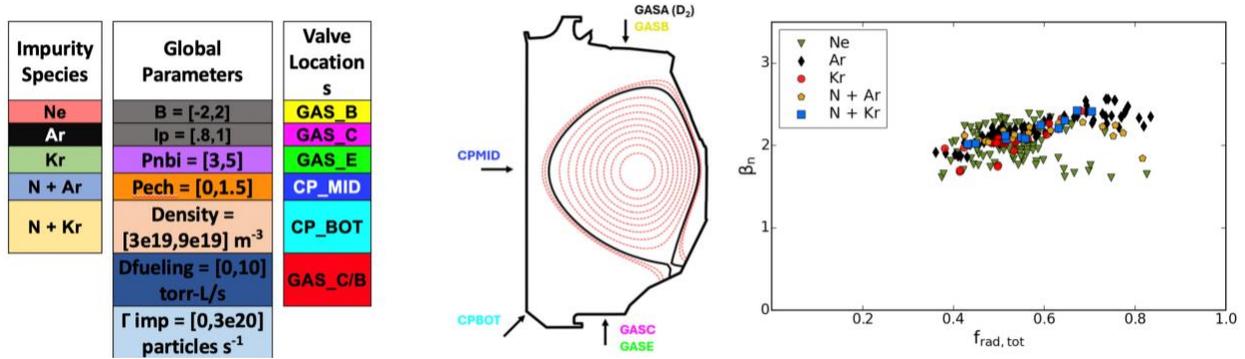

**Figure 1**. Left: table summarizing the achieved operational scenario, middle: cross section of the negative triangularity (NT) shape diverted discharges. The gas inlets used for seeding and fueling are also indicated. Right: Normalized plasma pressure as function of the total radiation fraction for the collected database with different seeding gas species (see labeled box).

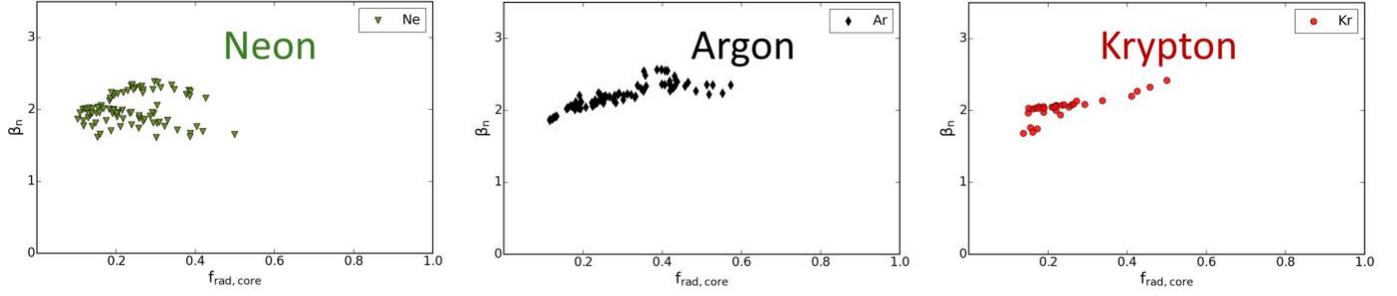

**Figure 2.** Normalized plasma pressure as function of the core radiation fraction for neon (green), argon (black) and krypton (red) seeded discharges.

## 2.1 Effect of impurity radiation on confinement and transport

Scenario integration with high performance and access to dissipative divertor conditions were assessed for all the available portfolio of extrinsic radiators (Ne, Ar, Kr, Ar+N, Kr +N) from low to high seeding rate and with both ion B×∇B drift into (fwd $B_T$) and out (rev $B_T$) of the lower divertor. The operational limits in terms of the maximum achievable radiation were explored and achieved for all radiators. An example of such limits is shown in figure 3 where we compare a discharge with a large amount of Kr (on the left, in blue), eventually leading to impurity accumulation and radiation collapse and a discharge where a reduced amount of Kr enabled the discharge to be sustained for >5s with stable performance.

During the experiments, we first puffed a high amount of Kr to find the operational limits (discharge #194306 in blue) and then the amount was lowered to obtain a stable discharge (discharge #194308 in green, on the right features a factor of 2 reduction in impurity concentration compared to #194306), while keeping all the other parameters constant. In panel a) –g) of figure 3 we show the main plasma parameters such as injected power, electron density, normalized beta, total radiated power, Kr seeded amount, Kr amount in the core as measured by the SPRED diagnostic and the neutron rate for the discharge on the left with high amount of Kr. Panels h)- n) on the right in green, show the same main parameters for the discharge with the lower amount of Kr. Improved performance are achieved as shown by the neutron rate, the plasma stored energy ($W_{mhd}$) and $\beta_N$ time traces.

The discharge with the large amount of Kr (figure 3 on the left with time traces in blue) is characterized by a steady increase in the total radiation through the discharge with a total radiation fraction of up 75%. The increase in the radiation is associated with an increase in plasma stored energy, normalized beta and electron density. The radiation increases rapidly and around 3.5sec exhibits a peak (plot d). This induces excess radiation over the injected power, the electron density increase eventually evolves into the Electron Cyclotron Heating (ECRH) cutoff leading to a radiation collapse, highlighting the critical role of the ECRH to avoid central impurity accumulation and calling for more ECRH availability in future experiments. For discharge #194308, featuring the lower level of Kr injection and a stable discharge evolution, the total radiation stays rather constant throughout the discharge and stays below 3MW, only towards the end there is a slight tendency to further increase, at t=4.8sec the discharge has reached $f_{rad}$ ~ 48% with stable performance. This enables the discharge to run smoothly till the programmed end where it can be ramped down as safely.

Interestingly, prior to the ECRH cutoff, discharge #194306 exhibits clear improved performance during the Kr high-seeded phase (red) compared to the lower Kr seeded phase (in blue). This is shown in the profile analysis (figure 4) which reveals a large increase in the ion temperature and plasma rotation at high radiation fraction (red profiles) compared to the lower Kr seeded phase (blue profiles). This demonstrates increase in plasma performance with increased radiation. The impurity injection broadens the profiles and increases the central value. The electron density profile is higher at the edge and reaches an almost 50% increase in the core. The increase in the electron density is caused mostly by steeping of the edge density profile. The ion temperature and carbon toroidal rotation rate have much broader profiles and higher central values as well. A large increase in the toroidal rotation is observed which together with an increase E×B shear reduce the turbulent particle flux as discussed later. The impurity injection yields to an increase of the electron density in the core creating a steeper electron density gradient (see figure 3 and figure 4). Similar effects are observed in discharge #194308. The confinement improvement during the high impurity seeded phase is also demonstrated by the increase of the plasma stored energy, $\beta_N$ as well as from the fact that the neutron rate does not degrade with the increased radiation (see figure 3). The main contribution to the neutron rate in DIII-D comes from the beam-target rate which is proportional to $T_e$ and $n_D/n_e$ with $n_D$ being the deuterium density [37]. The fact that no degradation of the neutron rate was observed, indicates that the increase in confinement during the seeded phase is larger than the dilution effect. These results are in agreement with studies using Kr injection in H-mode regime positive triangularity at AUG [38].

As the Kr content increases in the plasma and the ECRH availability is lost due to cutoff conditions, the discharge evolves towards core impurity accumulation and radiation collapse as demonstrated by the tomography reconstruction shown in figure 5. Kr accumulation leads to a strong radiative zone in the core where a pronounced poloidal asymmetry is measured on the low field side (figure 5 left). This centrifugal low field side localization of heavy impurity which arises from the fact that Kr with its high charge and mass can reduce the temperature screening and leads to impurity accumulation as also observed in standard positive triangularity plasmas [38, 39]. This highlights the importance of including such impurity transport effect which determine the distribution of the radiating ions in the neoclassical calculations.

To interpret the radiation measurements, the AURORA impurity transport code [29] has been applied to calculate the ionization equilibrium of the impurity species from the balance of sources and transport coefficients. This enables the inclusion of the effect of impurities on the local power balance through radiative cooling of the electrons, in addition to the fuel dilution that arises by imposing quasi-neutrality. In this work, we inferred the impurity density profiles and the associated impurity concentration using AURORA based on the constrains from VUV spectroscopy, bolometers and soft-X-ray [36, 40, 41]. Absolute density was constrained primarily by the radiation profile from bolometers, which has the most reliable absolute calibration and atomics data. The used atomics data are from the ADAS database [42]. On the other hand, SXR array and VUV spectrometer have a significantly better signal-to-noise-ratio and they helped to constrain shape and time evolution of the Kr density. In figure 6, we show the krypton radiation and density profile for the discharge with high Kr amount on the left and the one with lower Kr amount on the right at different times through the discharge.

While the Kr concentration associated with discharge #194308 (plots c) and d)) is well tolerated by the plasma till the ramp down, the impurity concentration associated with discharge #194306 leads to a rapid increase in the radiation and impurity density (figure 6 plots a) and b)). This induces impurity accumulation and collapse by radiation in agreement with results shown in figure 3 and 4. Also note, that the radiation and density profile near axis is flat or hollow due to on-axis ECH heating. Radiation collapse is preceded by a trip of ECH system due to a density cut-off, followed by on-axis accumulation as shown in Fig. 5a. Without the density cut-off, the plasma would likely survive even a higher Kr concentrations. This analysis is significant because, given the discharge condition analyzed here, provides the upper limit for impurity concentration which can be tolerated by those plasmas. Above such Kr level, impurity accumulation is triggered as a consequence of the reduction of the heat flux with high $P_{rad}$ and a reduction in the temperature screening, similar to results reported in [30]. In the conditions analyzed here, it is estimated that a Kr concentration of the order of $1\times10^{-4}$ is acceptable and can lead to confinement improvement before the onset of radiation collapse with an upper limit for the Kr concentration which lies around $2\times10^{-4}$.

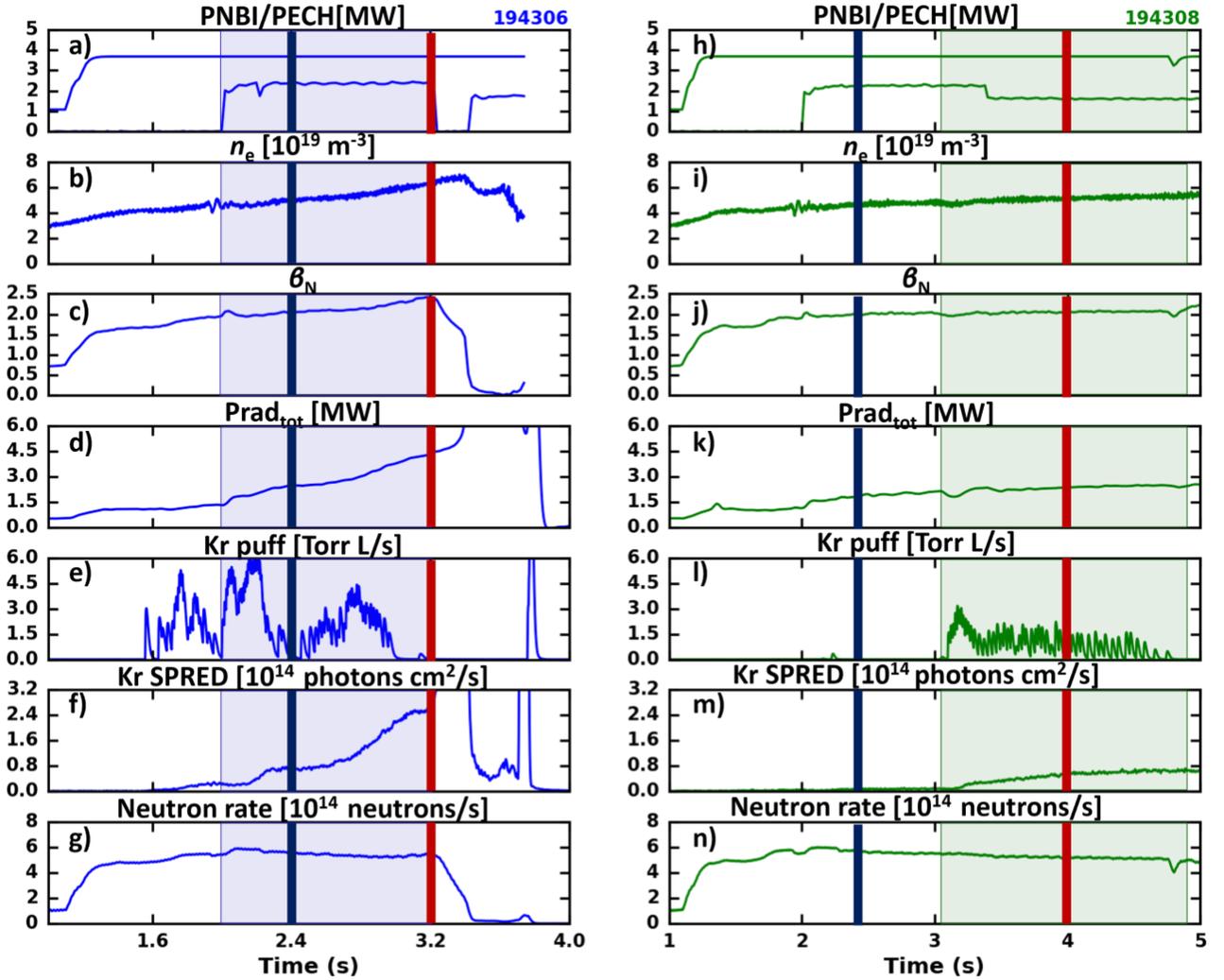

**Figure 3.** Time traces of main plasma parameters: a) injected power, b) electron density, c) normalized beta, d) total radiated power, e) Kr puff, f) Kr amount in the core as measured by the SPRED diagnostic, g) neutron rate for the discharge with high Kr seeding (#194306 in blue) where the high Kr amount leads to impurity accumulation ($f_{rad}$ = 75%). h) injected power, i) electron density, j) normalized beta, k) total radiated power, l) Kr puff, m) Kr amount in the core as measured by the SPRED diagnostic, n) neutron rate for discharge #194308 in green where the Kr concentration is reduced by a factor of 2 compared to the blue one. The vertical lines are the corresponding times of the profiles shown in figure 4.

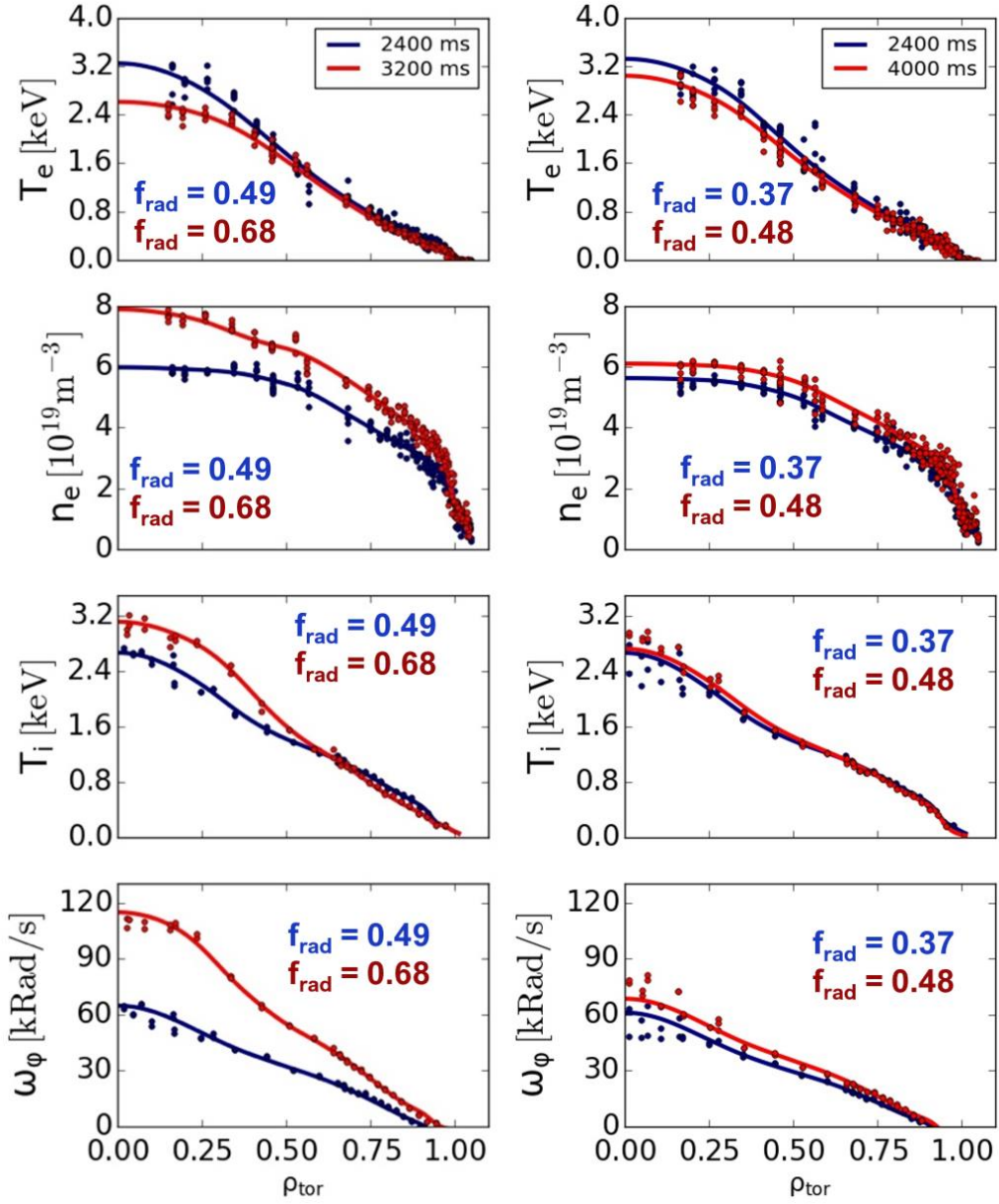

**Figure 4.** Left: $T_e$, $n_e$, $T_i$ and rotation profiles for the Kr seeded discharge #194306 at the beginning of the of the Kr injection (in blue, $f_{rad}$=0.49) and at high Kr level before impurity accumulation starts (red profiles, at higher Kr concentration, $f_{rad}$=0.68). Right: $T_e$, $n_e$, $T_i$ and rotation profiles for the Kr seeded discharge #194308 before the Kr injection (in blue, $f_{rad}$=0.37) and during a moderate level of Kr seeding (red profiles, $f_{rad}$=0.48). These profiles correspond to the vertical lines showed in figure 3.

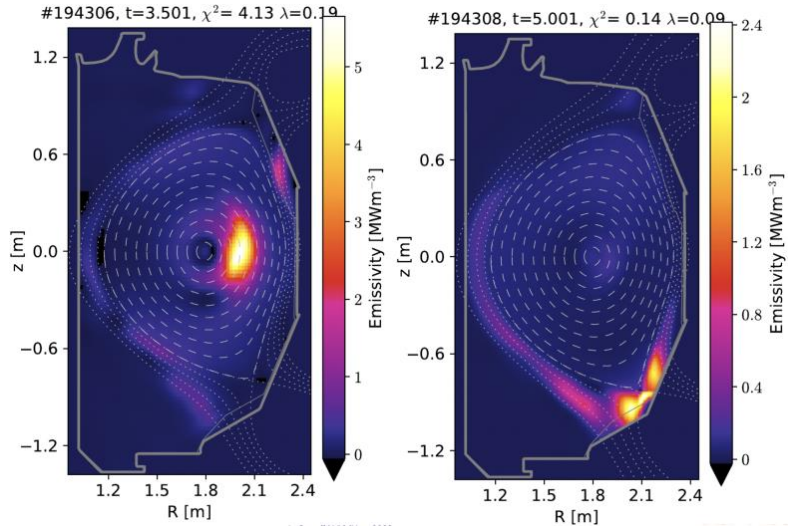

**Figure 5.** Tomography reconstruction of radiation profile of Kr seeded discharges from DIII-D foil bolometers. On the left, the large amount of Kr leads to core impurity accumulation and radiation collapse. Total radiation temporally exceeds the heating power and radiation at inner and outer strike-points vanished. Note the low field side poloidal asymmetry typical of heavy impurities in fast spinning plasmas. On the right: a Kr seeded discharge with moderate amount of Kr runs smoothly till ramp-down.

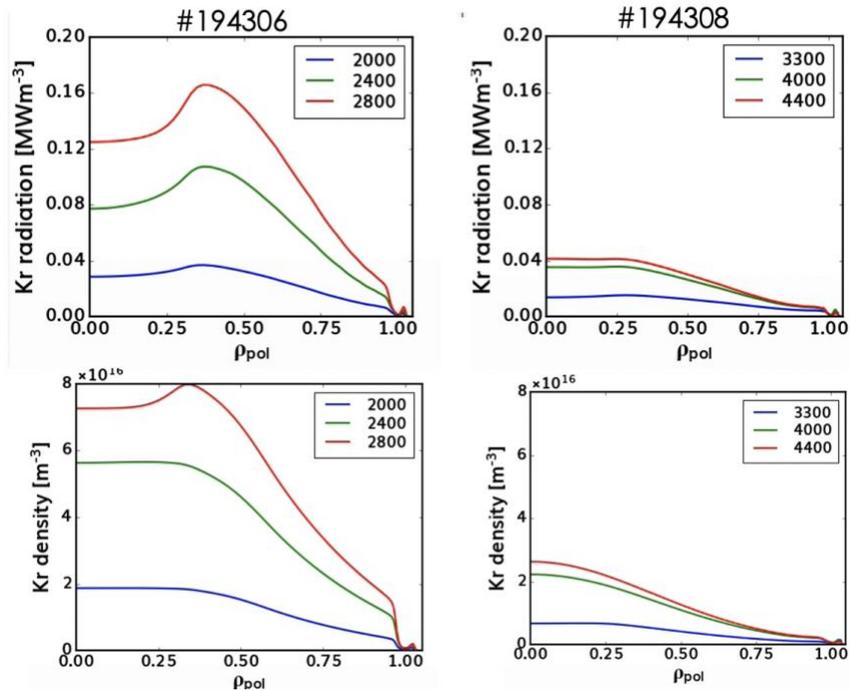

**Figure 6.** Krypton radiation (top) and density (bottom) as reconstructed by the AURORA code with constraints from experimental measurements of total and SXR radiation and $Kr^{25+}$ line observed by VUV spectrometer SPRED. Kr density increases in time due to a continual puff and high wall recycling. On the left for discharge #194306 which terminates due to impurity accumulation on the right for discharge #194308 with reduced Kr puff rate yielding a stable discharge that is safely ramped down.

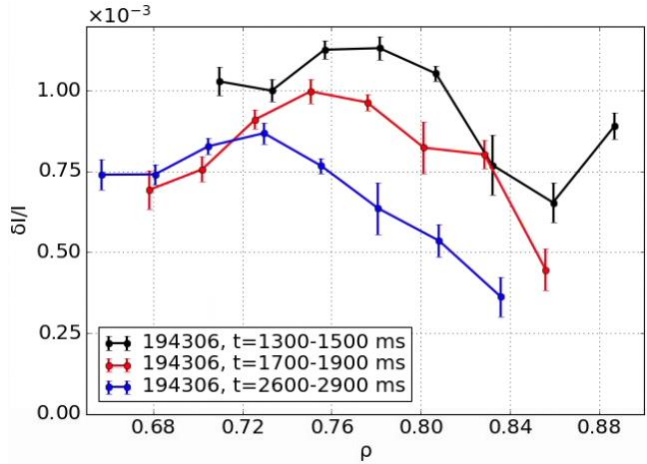

**Figure 7.** Amplitude of the density fluctuations in the frequency band $f = 60\text{-}280$ kHz as measured by the Beam Emission Spectroscopy (BES). The measurements reveal a reduction in long-wavelength density fluctuations over $0.6 < \rho < 0.85$ during impurity injection (red), impurity injection + ECRH (blue) compared to the unseeded phase (black).

These alterations in the radiation level and spatial distributions can be connected to the plasma confinement as discussed before. Therefore, we now investigate if related changes in plasma transport can be found. First, we inspect the density fluctuation levels. Detailed measurements of fluctuation ions of the intensity of the Doppler-shifted $D_\alpha$ light emission by the Beam Emission Spectroscopy (BES) [43] indicate changes to density fluctuations with impurity seeding. The amplitude of the density fluctuations in the frequency band $f = 60\text{-}280$ kHz is compared in figure 7 for a situation before impurity injection (black profiles) and during the seeding (red profiles) and during the impurity seeding + ECRH (blue profiles). These measurements show a reduction in the density fluctuations over $0.6 < \rho < 0.85$ during impurity injection (red and blue curves). As the impurity content increases, the fluctuation level decreases, which is connected to increased confinement and neutron rates until eventually a radiation collapse occurs. In order to understand this observation in the level of equivalent density fluctuations, we have conducted a kinetic transport analysis with the TRANSP code as well as TGYRO simulations where a reduction in particle, momentum, and ion heat is found which is consistent with the measurements. The analysis is presented in the next section.

## 3. TRANSP power balance analysis

Power balance analysis with the TRANSP [25] code has been performed using experimentally measured plasma profiles. We have used TRANSP in interpretative way such that the code inverts the transport equations to infer the transport necessary to match the observed evolution of fluid plasma profiles. In the power balance analysis is important to understand where the radiation losses are localized and how they affect the profiles. The analysis presented here allowed us to vary each key variable independently and directly identify the cause of the confinement improvement. This is important because in the experiments we cannot disentangle the cause and the consequence. Additionally, the TRANSP analysis enables to partially disentangle the simultaneous changes in $n_e$, $T_e$, $T_i$, heating, and fueling profiles and interpret them using more relevant quantities like heat momentum conductivity and particle diffusion. Results of this analysis for the Kr discharge are shown in figure 8. Similar results are found for argon and neon seeding discharges (fig. 9 and fig. 10) demonstrating that the same physics mechanism is at play for all those species. For all the three species, a decrease in both the electron and ion heat conductivity $\chi$ across much of the profile is found with increasing radiation fraction. The decrease is more pronounced in the ion channel where $\chi$ is reduced by a factor of 2 within $0.5 < \rho < 0.85$, see for instance fig. 8a. The reduction of the transport at mid-radius when a high fraction of the power is radiated is consistent with the location of the reduction in turbulence (fig. 7) needed to drive the heat flux. Similarly to the 50% decrease found in the ion conductivity, there is a reduction by a factor of 2 of the effective momentum diffusion (fig. 8b) resulting in higher toroidal rotation, which was also observed (see fig. 3). The reduction of the momentum diffusion is higher at larger radii, which hence further increases the toroidal flow shear inducing an increase in the resulting E×B shear (fig. 8d). This flow shearing and resulting E×B shearing is usually treated as a mean to reduce the turbulence growth rates and is therefore consistent with the reduction in equivalent density fluctuations measured by BES (figure 7). Higher source and reduced particle transport due to E×B shear increases density and collisionality (fig. 8g). The inverse electron density gradient scale length $R/L_{Te}$ (fig. 8e) remains mostly unchanged during the seeding while $R/L_{Ti}$ (fig. 8f) increases outside of mid radius.

Since the Kr behavior is determined by the evolution of the main plasma density profile through the inward convection, it is interesting to look closely at the behavior of the. inverse electron density gradient scale length $R/L_{ne}$ (fig. 8h). Largest density increase occurs close to pedestal (see density profile in fig. 3). The density profile at smaller radii is similar, increased by an offset which causes the decreases of $R/L_{ne}$ around $\rho = 0.4$ when Kr is injected. This indicates a more peaked density profile, which possibly is due to finite NBI fueling, reduced transport and the increase of the Ware pinch (neoclassical effect). This suggests that an influx of heavy impurities is taking place. Indeed, few ms later in the discharge, this impurity influx causes $R/L_{ne}$ to slightly decrease. Only near magnetics axis the density gradient increased and $R/L_{ne}$ returned to values close to these before impurity injection. Figure 8c) shows a non-negligible contribution of the radiation flux $Q_{rad}$ to the electron heat flux $Q_e$ not only at the very edge but also in the region from the edge towards mid-radius confirming the radiative mantle obtained with Kr. The "knee" observed in $Q_e$ ($\rho=0.2$) is due to the presence of the ECRH which is important to avoid accumulation of heavy impurities.

These results are in line with the Kr injection studies in standard H-mode found in [30]. Additionally, the study presented here in negative triangularity, shows similar results for krypton (figure 8), argon (figure 9) and neon (figure 10) seeded discharges highlighting that a similar physics mechanism is at play for the various impurity species.

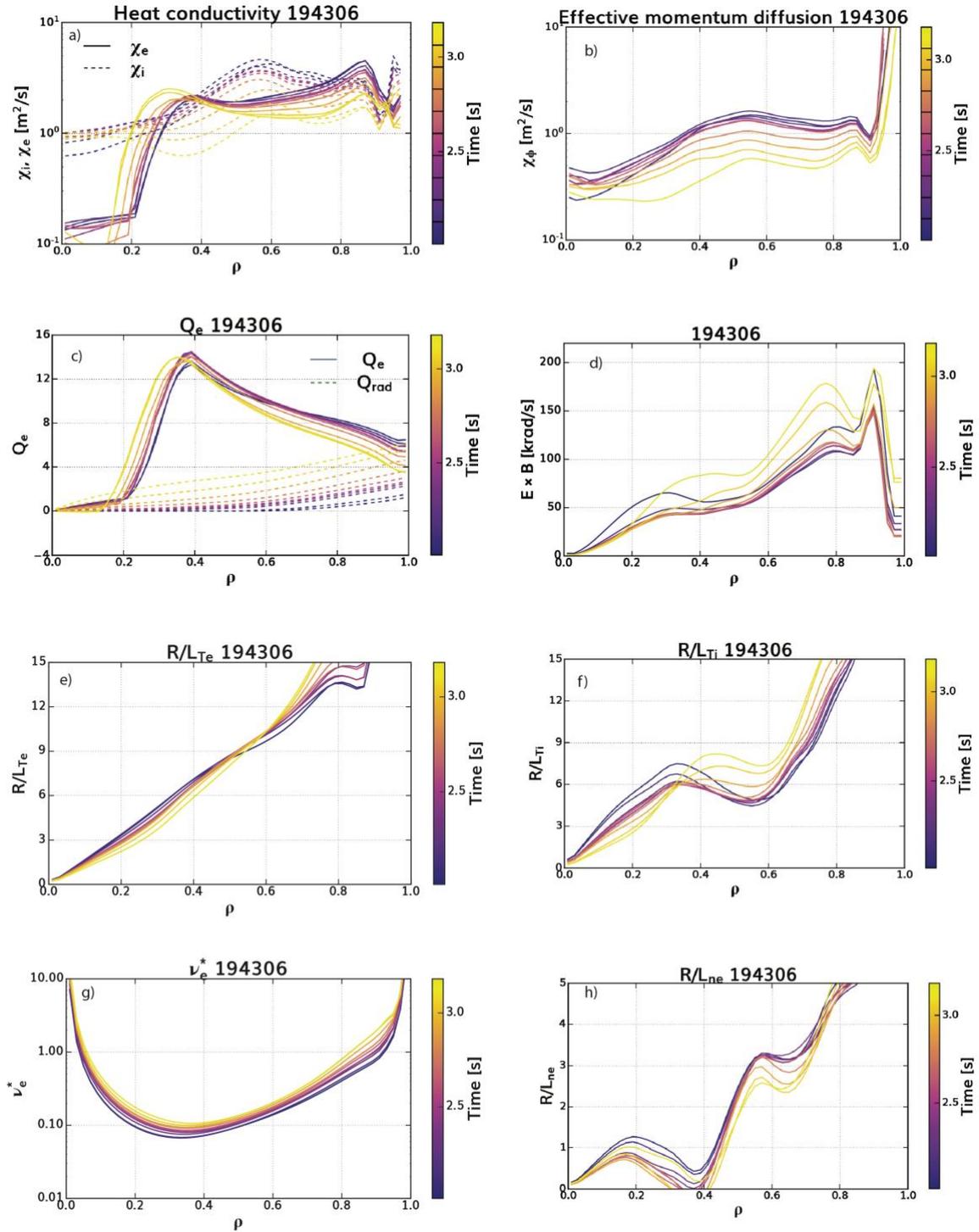

**Figure 8.** TRANSP analysis for a krypton seeded discharge showing profiles of a) ion and electrons heat conductivity, b) effective momentum diffusion, c) electron and radiation fluxes, d) E×B shear, e) $R/L_{Te}$, f) $R/L_{Ti}$, g) collisionality, h) $R/L_{ne}$. The different colors (from blue to yellow) represent the time evolution through the discharge.

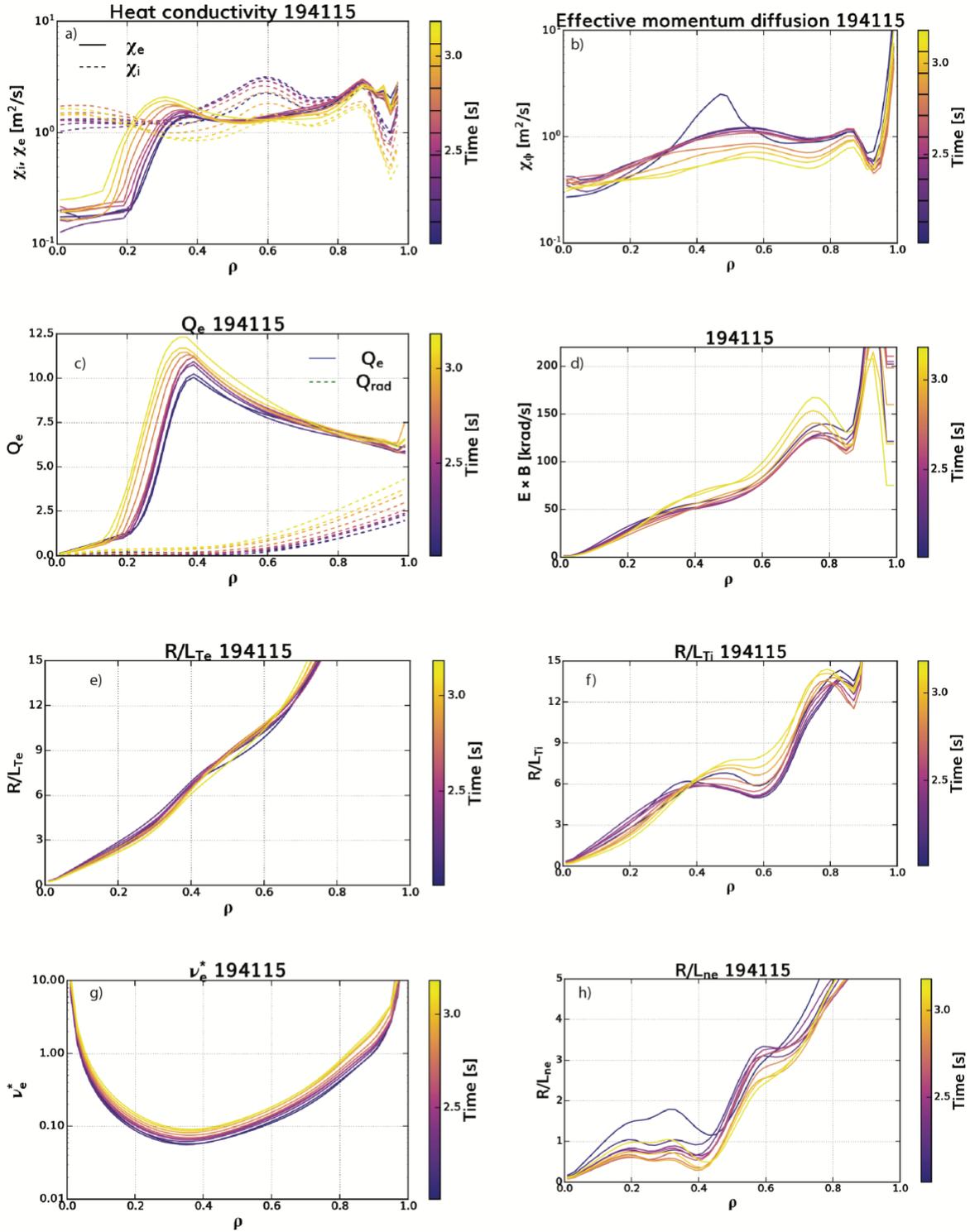

**Figure 9.** TRANSP analysis for an argon seeded discharge showing profiles of a) ion and electrons heat conductivity, b) effective momentum diffusion, c) electron and radiation fluxes, d) E×B shear, e) $R/L_{Te}$, f) $R/L_{Ti}$, g) collisionality, h) $R/L_{ne}$. The different colors (from blue to yellow) represent the time evolution through the discharge.

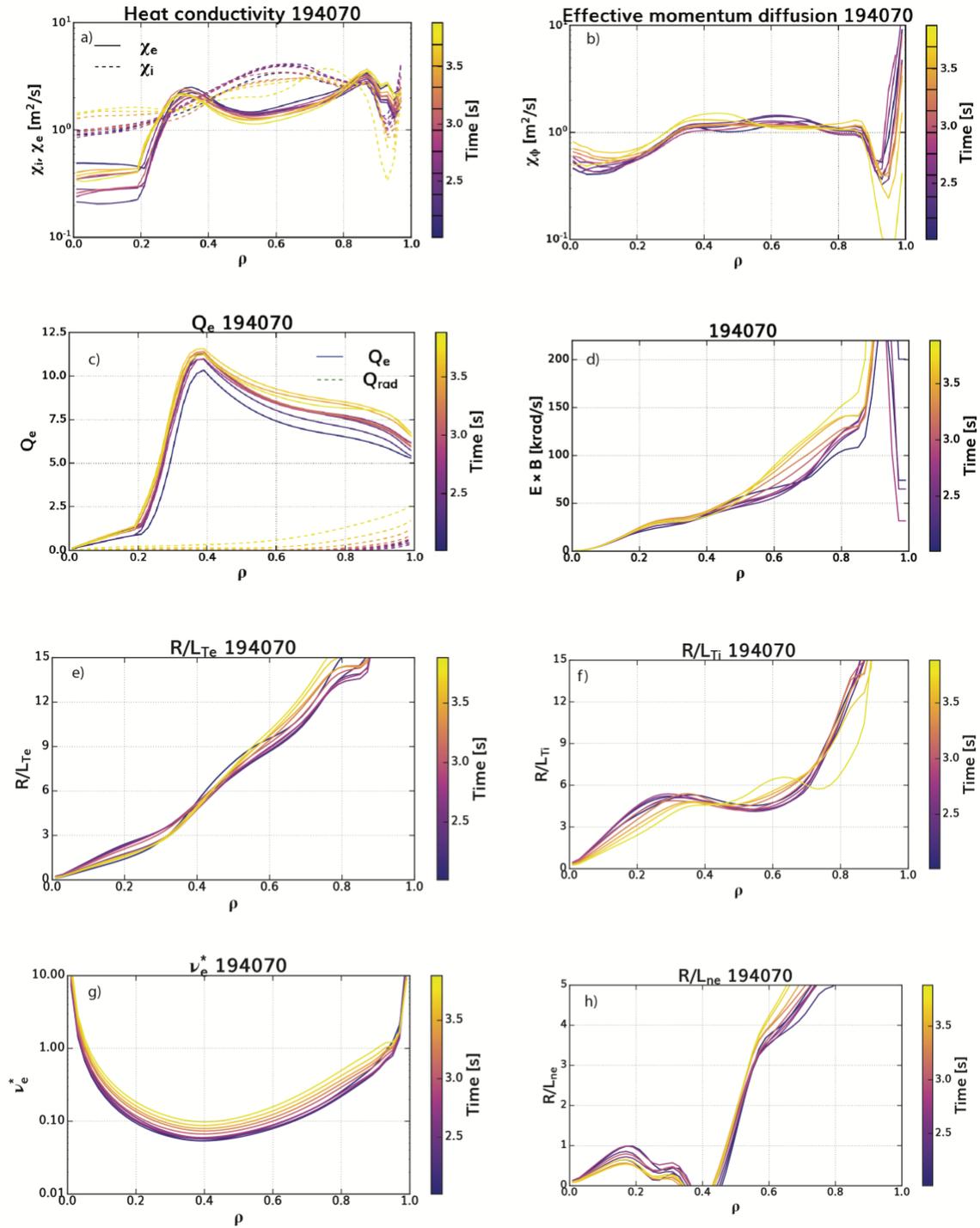

**Figure 10:** TRANSP analysis for a neon seeded discharge showing profiles of a) ion and electrons heat conductivity, b) effective momentum diffusion, c) electron and radiation fluxes, d) E×B shear, e) $R/L_{Te}$, f) $R/L_{Ti}$, g) collisionality, h) $R/L_{ne}$. The different colors (from blue to yellow) represent the time evolution through the discharge.

In order to quantify the effective outward transport of the impurities, we investigate effective particle diffusion (figure 11). We show this analysis for the argon discharge with similar results found for krypton and neon. The effective particle diffusion from TRANSP (figure 11 left) is defined as $\Gamma_e = D_{eff} \nabla n_e$ and it is decreases with time after the impurity injection (see color coded profiles over time). Figure 11 on the right shows impurity diffusion obtained from TGYRO (solid profiles) and NEO (dashed profiles) calculations in blue (no seeding) and in red (with seeding). These profiles show that the turbulence is reduced as the discharge progresses, while at the same time the neoclassical transport is increased later in the discharge, i.e. with higher impurity content. Note that neoclassical diffusion is increased by a higher collisionality (figure 8,9,10 plot ) and for heavy impurities also by a faster plasma rotation (figure3-4). These results suggest a mechanism where a reduction in the turbulent electron heat flux occurs as a consequence of the increased radiation losses. This reduces substantially the particle diffusion and particle source from NBI increases the electron density peaking which in turns increases the inward neoclassical pinch. This effect is particularly important for mid and high-Z such as Ar and Kr since the neoclassical effect increases with impurity charge Z [6]. The neoclassical flux is enhanced by a fast rotation of the plasma. Additionally, the temperature screening, which tends to counteract the inward convective term, gets smaller as well in fast spinning plasmas with increasing collisionality [38].

## 4. Profile prediction by TGYRO framework

The TGYRO framework [26] was used to predict a set of kinetics profiles matching the power balance fluxes from TRANSP analysis. This is achieved by iterative evaluation of TGLF and NEO, and identifying the best flux matching solution. In this study we used SAT2 rule in TGLF and $T_i$, $T_e$ and $n_e$ profiles were varies simultaneously with a fixed pivot at $\rho = 0.8$. Carbon density profile was fixed at experimental value and deuterium density was derived from the quasi-neutrality.

The radiative impurity was not included in the modeling directly as a species, but the measured impurity radiation was subtracted from the electron heat flux. This reduced the uncertainties in the calculation of radiated power by TGYRO. Figure 12 indicates that the kinetics profiles before the impurity injection are reasonably matched between experiment and modeling (left column of Fig. 12). However, after the impurity was injected, TGYRO was not able to explain the steep electron density profile (2$^{nd}$ column Fig. 12).

The density increase was so fast, that the actual power-balance particle flux was close to zero or negative. So, either there was a large inward pinch, unexplained by TGYRO, or a missing particle source from the seeded impurity. In the limit of infinitely slow impurity transport, the ionization and recombination of impurity are in balance. However, a finite radial transport causes a separation between ionization and recombination of the impurity and the impurity ionization will provide a finite electron source in the core even in the steady state case. Impurity recombination will result in an electron sink in the colder outer parts of the plasma. This source was calculated using the impurity transport code AURORA with the impurity transport coefficients inferred from the analysis presented above. Adding this ionization electron source to TGYRO modeling significantly improves the agreement with the measured density profile as shown in 3$^{rd}$ column of Fig. 12. The impurity ionization also causes an electron heat sink, however, it is negligible compared to the radiated power. The results highlight that impurity ionization in the core can be a significant particle source even in the steady state.

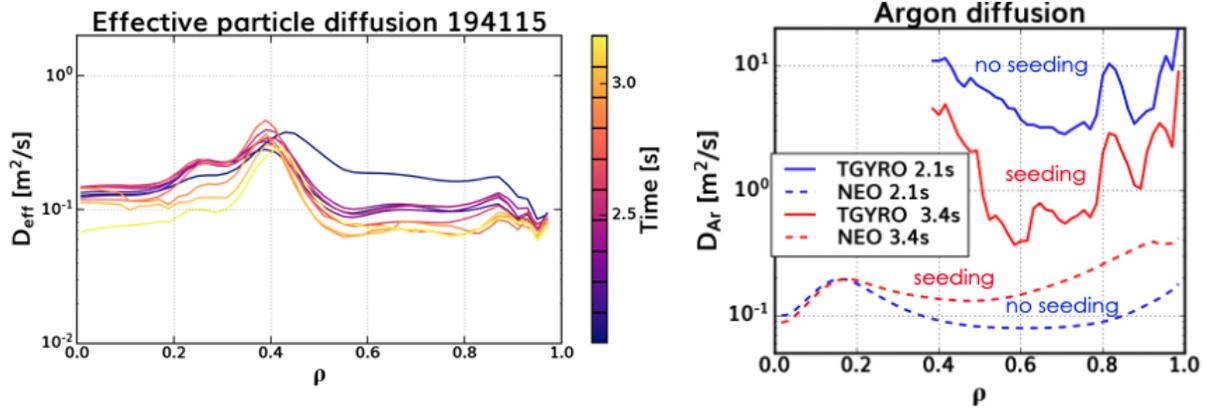

**Figure 11.** Effective particle diffusion from TRANSP (left). Argon diffusion from TGYRO and NEO calculations in blue at the beginning of the discharge (no seeding) and in red later in the discharge (seeding level).

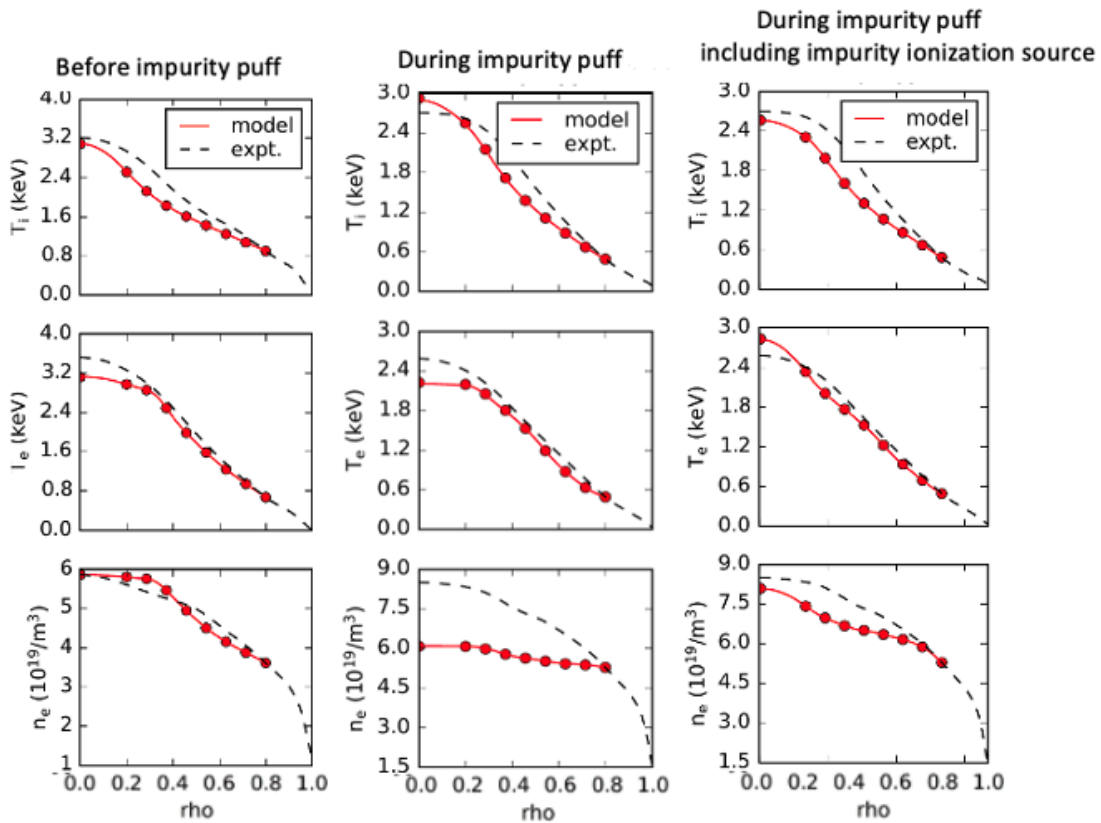

**Figure 12.** Kinetics profiles predicted by heat flux matching TGYRO simulation in the argon seeded discharge #194115.

## 5. Linear stability analysis

The output from the TGYRO analysis with matched heat flux was then transferred to TGLF [27] as inputs to perform parameter scans and linear stability analysis. The trapped-Gyro-Landau-Fluid (TGLF) is a reduced turbulent model to compute transport driven by drift-wave instabilities, such as Trapped Electron Modes (TEMs), Ion and Electron Temperature Gradient modes (ITG, ETG). The linear stability analysis shown here is important because it tells us what was the turbulent state inferred by TGYRO to explain the power balance fluxes.

Figure 13 illustrates a 2D map of the instability growth rate between the unseeded phase and the impurity seeded phase obtained through a TGLF scan. In this plot, the binormal wavenumber $k_y$ is plotted against radius with the signed growth rate as color coded quantity. These results indicates that when impurities are in the plasma (Fig. 13 on the right), the low-$k_y$ ion dominated region of growth rate (blue) expands compared to the unseeded case (Fig. 13 on the left), whereas the electron-dominated region (red) shrinks. This might be because the steeper ion temperature profiles presented in the earlier analysis (fig. 4), destabilizing the ITG turbulence, while the higher collisionality (fig. 8, 9,10 plots) caused by increased ne and reduced $T_e$, reduces the grow rate of TEM.

TGLF scan was performed, with the heat flux matched solution at one single location. In figure 14 the mid-radius ion scale growth rate is shown and compared to the growth rate at the two characteristic time points early in the discharge (no seeding) and later on (with seeding). In fig. 14, the horizontal line represents E×B shear with full symbols representing dominant ion drift direction modes and open symbols electron drift direction modes. Although the low-k growth-rate increases with impurity injection, as shown in both figures 13 and 14, the E×B shear rate increases even more and approaches the linear growth rate of the turbulence. Therefore, E×B shear might be the main mechanism responsible for turbulence stabilization and observed improvement in plasma confinement.

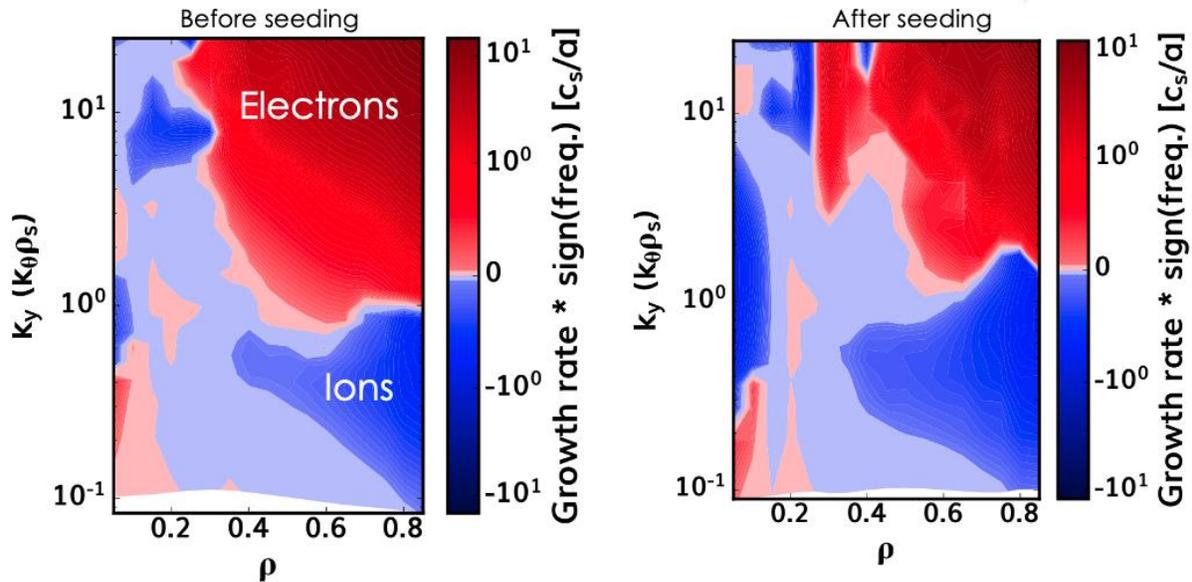

**Figure 13.** Growth rate profiles from TGLF showing the expansion of the low-$k_y$ ion dominated region and the reduction of the electron-dominated region before impurity seeding (left) and during impurity seeding (right).

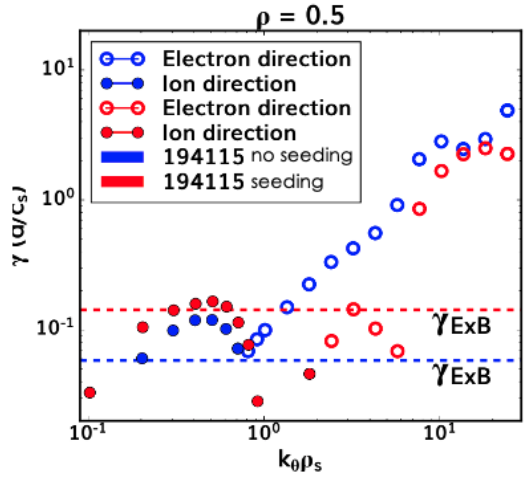

**Figure 14.** $k_y$ spectrum in discharge 194115 before seeding (2.1s blue) and during (3.4s red). Dashed horizontal line represents E×B shearing rate.

## 5.1 Parameter sensitivity scans by TGLF code

In order to investigate the parameter dependence of the turbulence transport and gain insight in the underlying physics of the transport stabilization, extensive 2D parameter sensitivity scans were performed using TGLF. Scans were performed starting from experimentally measured profiles and gradients in the argon seeded discharge 194115 at $\rho = 0.5$. The most important question is how the fluxes depend on E×B shear, $Z_{eff}$, triangularity $\delta$, collisionality $\upsilon$ and $T_e/T_i$, because all these parameters were affected by impurity seeding. Here we have used the plasma background of the unseeded phase and scan parameters towards the seeded case. The Exp value is the experimental value of the flux from power balance.

First scan is presented in Fig. 15 and demonstrates a strong reduction of heat fluxes (Q) and momentum flux ($\Pi$) with electron collisionality. $Q_i$ and $\Pi_i$ also decreases with increasing $T_i/T_e$ ratio and outward particle flux ($\Gamma$) moderately increases with collisionality. The percent in the plots of figure 15 represent the relative difference compared to TGYRO calculated flux values.

Effect of local triangularity, investigated in Fig. 16 is significantly weaker. Triangularity has not changed by the seeding, but we are interested how the fluxes change for positive value. Heat fluxes are changing just by a few percent and the same for the ion momentum flux. The electron particle flux is affected the most and it becomes more inward. The most significant triangularity effects originates from the edge of the plasma and therefore they are not captured be these local parameter scans on mid-radius. This also supports the finding that the underlying physics of impurity induced confinement and transport stabilization occurs in both PT and NT configurations.

The last two essential quantities are E×B and $Z_{eff}$. Their 2D scan is presented in Fig. 17. Increase in E×B strongly reduces all fluxes. Because E×B depends on the plasma rotation, the reduction of momentum flux from NBI increases the E×B shear and amplifies the stabilization effect. $Z_{eff}$ dependence is significantly weaker and it slightly reduces the electron heat flux and increases particle flux. The other parameters like magnetics shear can also reduce turbulence growth-rate, however it was not significantly affected by impurity seeding.

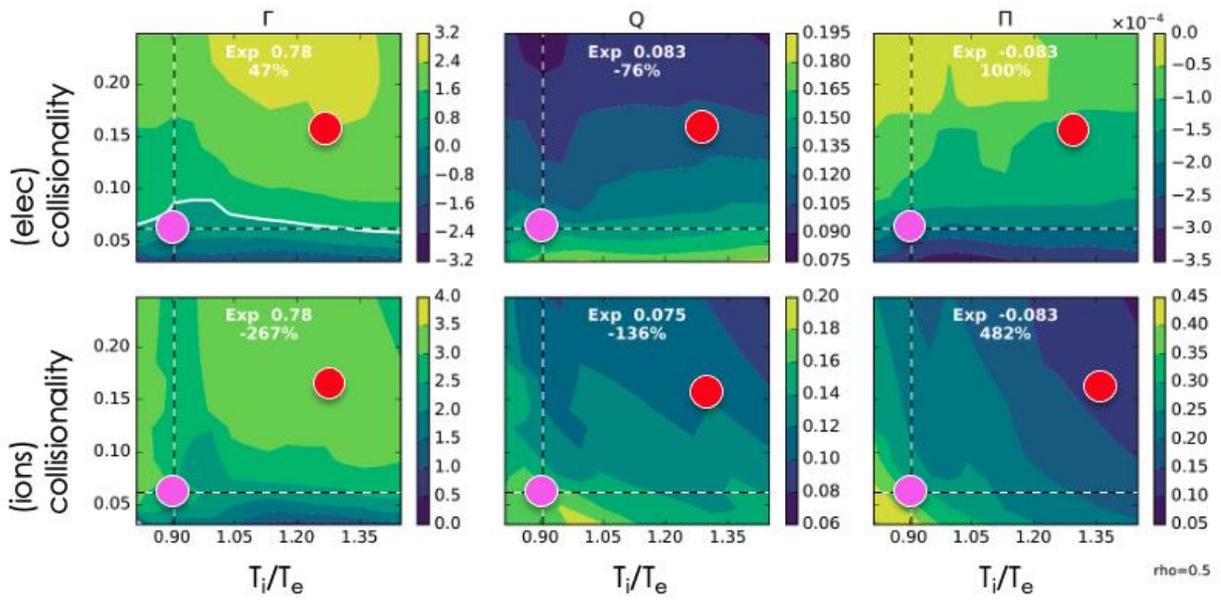

**Figure 15.** 2D scan of particle flux Γ, heat flux Q and momentum flux Π versus electron collisionality and $T_i/T_e$ ratio. Magenta dots represent conditions before argon seeding, red dots during the seeding.

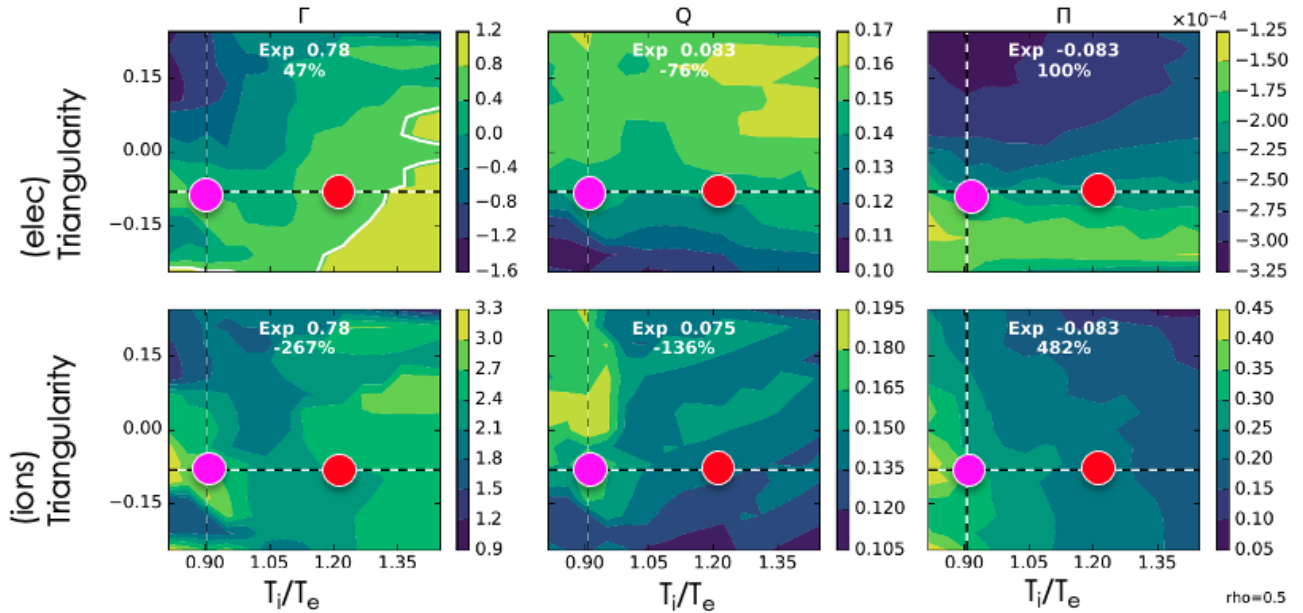

**Figure 16.** 2D scan of particle flux Γ, heat flux Q and momentum flux Π versus triangularity and $T_i/T_e$ ratio. Magenta dots represent conditions before argon seeding, red dots during the seeding.

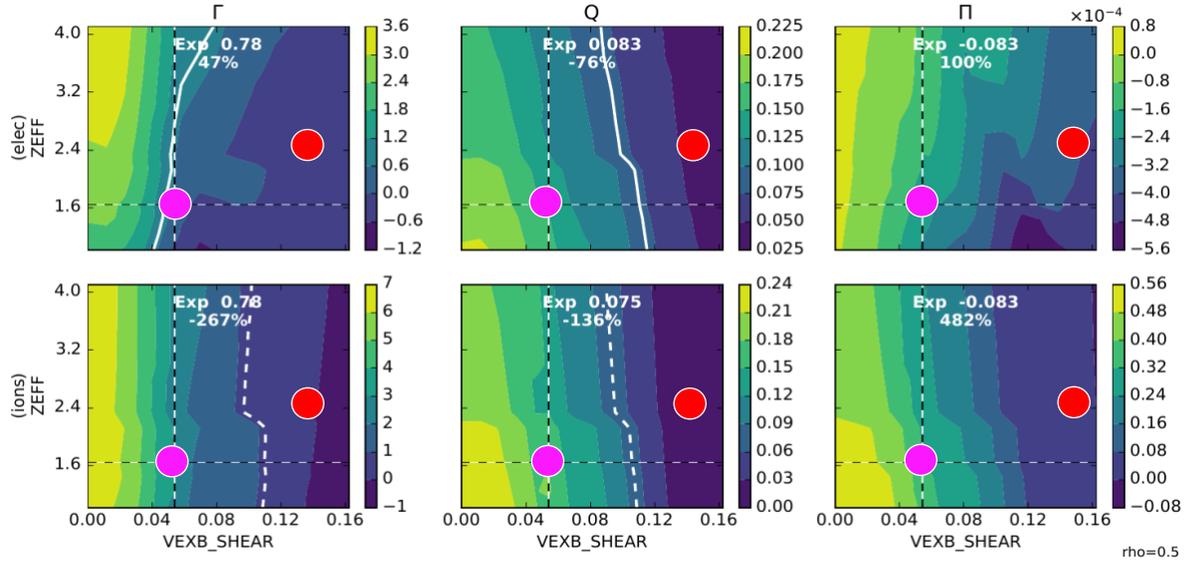

**Figure 17.** 2D scan of particle flux Γ, heat flux Q and momentum flux Π versus triangularity E×B and $Z_{eff}$. Magenta dots represent conditions before argon seeding, red dots during the seeding.

## 6. Discussion and Conclusion

This work presents the first achievement of highly radiating plasmas in strongly shaped negative triangularity. We have collected a comprehensive database which spans a wide parameter range with a total radiated power fraction up to 0.85. Mantle radiation limits have been explored and achieved with neon, argon and krypton seeding. Nitrogen seeding has been added in some discharges as extrinsic divertor radiator in addition to carbon (which is an intrinsic impurity at DIII-D) to trigger access to detachment conditions. Reactor relevant values of normalized plasma pressure $\beta_N$ ($\beta_N > 2$) were achieved with krypton and argon at high radiation fraction and high Greenwald fraction.

The findings presented here demonstrate how impurity seeding can lead to confinement improvement in strongly shaped ELM free negative triangularity plasmas. We have shown that the injection of krypton, argon and neon affected core plasma in two different ways. First the impurity radiation losses reduce electron heat flux $Q_e$, electron temperature $T_e$, and $T_e$ gradient. Lower $T_i/T_e$ and $R/L_{Te}$ reduces drive for TEM which reduces the momentum transport (see Fig. 15). Moreover, lower $T_e$ increases collisionality which also reduces the TEM growth rate. Reduced momentum transport and finite momentum source from beams increases the plasma rotation and E×B shear which approaches the low-k modes growth rate (see Fig 14) and reduces all fluxes as shown in Fig. 17, including the momentum flux. Lower momentum flux increases plasma rotation even further and this amplifies the observed effects. The second effect of the impurities is the increased core electron particle source. Higher source and reduced particle transport due to E×B shear increases density and collisionality. Higher density increases plasma radiation, which is proportional to $n_e$ and $n_{imp}$ and higher collisionality is reducing TEM growth rate, heat fluxed and momentum flux, as mentioned earlier. Shrinking of electron dominated spectra in Fig. 11 confirms that the TEM is suppressed and ITG becomes a prevalent plasma instability. This mechanism raises the question of how the plasma will respond in device without significant external momentum and particle input from neutral beams. The confinement improvement might be less prevalent in low collisionality FPP plasmas, because the small increase in collisionality will not affect TEM. Our future experiments will focus on radiation effects in DIII-D plasma in PT and NT with balanced beams or with a pure wave heating by ECH. These results are in line with impurity seeding enhancing performance observed in positive triangularity [44, 45, 9], thus demonstrating that a robust physics mechanism is at play.

These new results on the effect of impurity seeding on confinement and transport in highly radiative NT plasmas demonstrate that 1) the physics mechanism governing impurity inducing confinement improvement involves impurity stabilization through dilution as well as profiles changes due to impurity radiation cooling, 2) this physics mechanism is very robust and appears to hold for a variety of impurity portfolio and is independent on the plasma configuration including positive and negative triangularity. It has to be noted that in the NT studies presented here the increase in confinement triggered by impurities is in addition to the increase in confinement typical of NT configuration due to the reduced turbulence transport in the core region and consequence of the weaking of the TEM instabilities by the NT shape [23]. Thus, the shift from TEM to ITG due to the impurity seeding effect may appear stronger in NT compared to PT as the impurity effect builds up on different starting conditions with a more pronounced density peaking and a different dominant mode.

Another important finding is that the impurity seeded NT discharges exhibit a strong electron density peaking (with the term "peaking" we refer to the ratio of the density at rho=0.2 and the volume-averaged density) which in general appears to not be associated with impurity accumulation. It remains important to maintain the impurity level in the core low to avoid excessive fuel dilution. This is more challenging with neon due to its high ionization potential and radiation characteristics [11, 30] compared to higher-Z impurities such as argon and krypton which have higher radiative efficiency and lower ionization potential. Both the high ionization potential and radiative efficiency of neon in the considered $T_e$ range, can quickly lead to high fuel dilution potentially resulting in an overall confinement degradation. With that in mind, as long as the impurity level in the core is kept low to avoid excessive fuel dilution and impurity accumulation, integration of NT configuration with high radiation by mid and high-Z impurities not only is achievable as demonstrated by these first experiments but it can be also advantageous. The advantages are that 1) higher core radiation is allowed in NT due to the absence of the requirement to stay in H-mode, effectively mitigating the power exhaust problem, 2) impurity confinement in NT is such to indicate a favorable impurity transport and 3) a mechanism of impurity induced confinement improvement is at play. The impurity effect is a combination of local transport changes and density profiles variation. To this end, SOL physics and its effect on such variations becomes important. The study of the effect of impurities on SOL and divertor physics in these highly radiative experiments employing reactor-relevant seeding is described in another publication [30].

The finding presented here are a significant step towards assessing the viability of the negative triangularity path NT as a potential reactor operational regime. Highly radiating NT plasmas offer a unique opportunity to integrate a radiative scenario with high core performance and intrinsically no ELMs.

**Acknowledgments**

This material is based upon work supported by the U.S. Department of Energy, Office of Science, Office of Fusion Energy Sciences, using the DIII-D National Fusion Facility, a DOE Office of Science user facility, under Award(s) DE-SC0023100, DE-FC02-04ER54698, DE-SC0022270, DE-AC52-07NA27344, DE-FG02-08ER54999.